\documentclass[aps,tightenlines,nofootinbib,11pt,longbibliography,superscriptaddressm,notitlepage]{revtex4-1}
\usepackage{graphicx}  
\usepackage{dcolumn}   
\usepackage{bm}        
\usepackage{amssymb}   
\usepackage{xcolor}
\usepackage[normalem]{ulem}
\usepackage{amsmath}
\usepackage{mathtools}
\usepackage[english]{babel}
\usepackage{hyphenat}
\usepackage{accents}
\usepackage{natbib}
\usepackage{enumitem}
\usepackage[bottom]{footmisc}
\usepackage[section]{placeins}
\usepackage{dsfont}
\usepackage{natbib}
\usepackage{relsize}
\usepackage{setspace}
\usepackage{etoolbox}
\usepackage{graphicx}
\usepackage[caption=false]{subfig}

\patchcmd{\section}
  {\centering}
  {\raggedright}
  {}
  {}
\patchcmd{\subsection}
  {\centering}
  {\raggedright}
  {}
  {}
  
  \usepackage{mciteplus}
    \usepackage{hyperref}

\newcommand{\de}{\mbox{d}}
\newcommand{\e}{\mbox{e}}
\newcommand{\sscr}{\scriptscriptstyle}
\newcommand{\be}{\begin{equation}}
\newcommand{\ee}{\end{equation}}

\newcommand{\pa}{\partial}

\newcommand{\GFTbounce}{Oriti:2016qtz,*Oriti:2016ueo}
\newcommand{\Markov}{markov1982limiting,*markov1987possible}
\newcommand{\Frolov}{Frolov:1988vj,*Frolov:1989pf}

\mciteSetSublistMode{f}

\numberwithin{equation}{section}

\begin{document}

\title{Reconstruction of mimetic gravity in a non-singular bouncing universe from quantum gravity}

\author{Marco de Cesare}
\email{marco.de\_cesare@unb.ca}
\affiliation{Department of Mathematics and Statistics, University of New Brunswick, Fredericton, NB, Canada E3B 5A3}

  \begin{abstract}
We illustrate a general reconstruction procedure for mimetic gravity. Focusing on a bouncing cosmological background, we derive general properties that must be satisfied by the function $f(\Box\phi)$ implementing the limiting curvature hypothesis. We show how relevant physical information can be extracted from power law expansions of $f$ in different regimes, corresponding e.g.~to the very early universe or to late times. Our results are then applied to two specific models reproducing the cosmological background dynamics obtained in group field theory and in loop quantum cosmology, and we discuss the possibility of using this framework as providing an effective field theory description of quantum gravity. We study the evolution of anisotropies near the bounce, and discuss instabilities  of scalar perturbations. Furthermore, we discuss two equivalent formulations of mimetic gravity: one in terms of an effective fluid with exotic properties, the other featuring {\it two} distinct time-varying gravitational ``constants'' in the cosmological equations.
  \end{abstract}
  
\maketitle

\section{Introduction}
The resolution of spacetime singularities is one of the main expected consequences of quantum gravity. 
In cosmology, the realisation of such a possibility would lead to the replacement of the Big Bang singularity by a smooth spacetime region, e.g.~a bounce, with profound implications for our understanding of the earliest stages of cosmic expansion and of the initial conditions for our Universe. Non-singular bouncing cosmologies have been extensively studied and may represent an alternative to the inflationary scenario \cite{Brandenberger:2016vhg} {with specific observational signatures (see also \cite{Cai:2014bea}).}
Resolution of the initial singularity in cosmology has been achieved in various approaches based on a loop quantisation of the gravitational field, such as loop quantum cosmology (LQC)~\cite{Bojowald:2008zzb,Ashtekar:2011ni}, group field theory (GFT) condensate cosmology \cite{\GFTbounce}, and quantum reduced loop gravity \cite{Alesci:2016xqa}; more specifically, both in LQC and in GFT the initial singularity is replaced by a regular bounce, marking the transition from a contracting phase to an expanding one.

One of the main open problems, that is common to all background-independent approaches to quantum gravity, is the derivation of an effective field theory taking into account effects due to the underlying discreteness of spacetime at the Planck scale.
In fact, at present very little is known about quantum gravity beyond perfect homogeneity, although efforts to include inhomogeneities in the description of an emergent universe from full quantum gravity are underway \cite{Gielen:2017eco,Gerhardt:2018byq,Gielen:2018xph}.
One possible alternative approach then consists in considering modifications of general relativity that are able to reproduce known features of a given quantum gravity theory. The hope is that, by doing so, we can gain insight (at least qualitatively) into the consequences of quantum gravitational effects in different regimes.
In this work, we adopt the framework of limiting curvature mimetic gravity and examine in detail the problem of  reconstructing the theory from the evolution of the cosmological background, with particular attention to the case of a bouncing background.  Such a theory should then be regarded as a toy model for an effective description of quantum gravity \cite{Bodendorfer:2017bjt,Liu:2017puc} {and can be used to study its phenomenological consequences.} Possible applications include e.g.~the dynamics of inhomogeneous and anisotropic degrees of freedom in cosmology, and black holes.

The idea of limiting curvature as a possible solution to the singularities of  general relativity was first envisaged in Ref.~\cite{\Markov,\Frolov}, and subsequently implemented in modifications of the Einstein-Hilbert action including higher-order curvature invariants in Refs.~\cite{Mukhanov:1991zn,Brandenberger:1993ef,Easson:1999xw,Yoshida:2017swb}. An alternative proposal for constructing a gravitational theory with a built-in limiting curvature scale was put forward in Ref.~\cite{Chamseddine:2016uef} as an extension of mimetic gravity. This is achieved by including in the action functional  a (multivalued) potential term $f$ depending on the d'Alembertian of a scalar field $\phi$. Upon closer inspection, such a potential turns out to depend on the expansion scalar $\chi$ of a privileged irrotational congruence of time-like geodesics, singled out by the so-called mimetic constraint \cite{deCesare:2018cts}. On a cosmological spacetime, $f(\chi)$ reduces to a function of the Hubble rate \cite{Langlois:2018jdg}. Multivaluedness of the potential is necessary for a consistent realisation of bouncing cosmologies in this framework \cite{Brahma:2018dwx,deHaro:2018sqw,deHaro:2018cpl,deCesare:2018cts}. Non-singular black hole solutions have been studied in Refs.~\cite{Chamseddine:2016ktu,BenAchour:2017ivq}.

The particular model proposed in Ref.~\cite{Chamseddine:2016uef} exactly reproduces the effective dynamics obtained in (flat, isotropic) homogeneous loop quantum cosmology. Thus, all curvature invariants are bounded throughout spacetime by a limiting curvature scale, which is in turn related to the existence of a critical value for the energy density of matter at the bounce. From the point of view of quantum gravity, it is natural to require that the limiting curvature scale be Planckian. In Ref.~\cite{Liu:2017puc} a broader class of theories was identified in the DHOST family, all reproducing the effective dynamics of loop quantum cosmology; these models can be further extended by the inclusion of a term corresponding to the spatial curvature.
The relation between the model of Ref.~\cite{Chamseddine:2016uef} and effective loop quantum cosmology was further investigated in Refs.~\cite{Bodendorfer:2017bjt,Bodendorfer:2018ptp} from a Hamiltonian perspective, showing that the equivalence holds in the spatially flat, homogeneous and isotropic sector; however, the correspondence is lost in the anisotropic case. Nevertheless, even for anisotropic cosmologies the solutions of the two models are qualitatively similar~\cite{Bodendorfer:2017bjt,Bodendorfer:2018ptp}.
The mimetic model of Ref.~\cite{Chamseddine:2016uef} has been recently generalised in Ref.~\cite{deCesare:2018cts}, where a limiting curvature mimetic gravity theory was reconstructed so as to exactly reproduce the background evolution obtained from group field theory condensates in Ref.~\cite{\GFTbounce}; the effective dynamics of homogeneous loop quantum cosmology is then recovered as a particular case for some specific choice of the parameters of the model.

This paper has two main goals. The first one is to give a general account of theory reconstruction in mimetic gravity, showing how essential information about background evolution (e.g.~the critical energy density, the bounce duration, and the equation of state of effective fluids) is encoded in the function $f(\chi)$, particularly in its asymptotic behaviour in regimes of physical interest.
The case of a generic bouncing background is examined in detail, although our methods have a much broader applicability.
We provide general prescriptions for the matching of the different branches of the multi-valued function $f(\chi)$, which are necessary in order to obtain a smooth evolution of the universe, thus generalising the analysis of matching conditions in Ref.~\cite{deCesare:2018cts}.
{Our second goal is to study in detail the properties of mimetic gravity theories with the same background evolution as obtained in non-perturbative approaches to quantum gravity. Specifically,} we analyse the model of Ref.~\cite{deCesare:2018cts} reproducing the background evolution obtained from GFT condensates, and compare it to the special case corresponding to the LQC effective dynamics. 
We study the evolution of anisotropies near the bounce in a Bianchi~I spacetime, including the effects of hydrodynamic matter with generic equation of state, thus extending the results of Ref.~\cite{Chamseddine:2016uef}. As in the model of Ref.~\cite{Chamseddine:2016uef}, our more general results also show that the smooth bounce is not spoiled by anisotropies, which stay bounded during the bounce era. Instabilities in the inhomogeneous sector are also discussed. Moreover, given its relevance and simplicity, the particular case corresponding to the effective dynamics of LQC is analysed separately.

The plan of the paper is as  follows.
The formulation of mimetic gravity is briefly reviewed in Section~\ref{Sec1}. In Section~\ref{Sec:Reconstruction} we discuss the reconstruction procedure. In Section~\ref{EffectiveGFT} we focus on the model of Ref.~\cite{deCesare:2018cts}: we discuss the background evolution, exhibit the form of the function $f(\chi)$ and derive
its expansion in its two branches, corresponding to the region around the bounce and to a large universe. The model of Ref.~\cite{deCesare:2018cts}, which can be obtained as a particular case from our more general model, is discuss separately due to its relevance and simplicity. Section~\ref{Anisotropies} is devoted to the study of anisotropies in a bouncing background.  In Section~\ref{GravConst} we provide an alternative description of the cosmological dynamics of mimetic gravity in terms of two effective gravitational ``constants'', both depending on the expansion rate of the universe. In Section~\ref{Insta} we discuss instabilities of scalar perturbations. We conclude with a discussion of our results in Section~\ref{Conclusions}.

We choose units such that $8\pi G=1$. Landau-Lifshitz conventions for the metric signature $(+---)$ are adopted.

\section{Mimetic gravity and its cosmology}\label{Sec1}
The version of mimetic gravity considered in Ref.~\cite{Chamseddine:2016uef} is based on the action
\be\label{ActionPrinciple}
S[g_{\mu\nu},\phi,\lambda,\psi]=\int\de^4x \sqrt{-g}\; \left(-\frac{1}{2}R+\lambda(g^{\mu\nu}\pa_\mu\phi\pa_\nu\phi-1)+f(\chi)+L_{\rm m}(\psi,g_{\mu\nu})\right) ~,
\ee
with $\chi=\Box\phi$.
The gravitational sector consists of the metric $g_{\mu\nu}$ and the scalar field $\phi$. The Lagrange multiplier $\lambda$ enforces the mimetic constraint
\be\label{MimeticConstraint}
g^{\mu\nu}\pa_\mu\phi\pa_\nu\phi=1~.
\ee
We have included a matter Lagrangian $L_{\rm m}$, where $\psi$ represents a generic matter field, coupled to $g_{\mu\nu}$ only and not to $\phi$.
Due to the term $f(\chi)$, the action (\ref{ActionPrinciple}) represents a higher-derivative extension of the original mimetic gravity theory of Ref.~\cite{Chamseddine:2013kea}.\footnote{The original formulation of mimetic gravity of Ref.~\cite{Chamseddine:2013kea} relied on a singular disformal transformation \cite{Deruelle:2014zza} (see also Ref.~\cite{Langlois:2018jdg}). An equivalent formulation with a Lagrange multiplier implementing the constraint (\ref{MimeticConstraint}) was given in Ref.~\cite{Golovnev:2013jxa}. The latter represents the starting point for further generalisations of the model considered in Refs.~\cite{Chamseddine:2014vna,Chamseddine:2016uef}. See also the review~\cite{Sebastiani:2016ras}.}

Due to the mimetic constraint, the vector field $u^\mu=g^{\mu\nu}\pa_{\nu}\phi$ has unit norm and generates an irrotational congruence of timelike geodesics (see Ref.~\cite{deCesare:2018cts} for more details). Thus, the theory admits a preferred foliation\footnote{Such a gauge choice corresponds to unit lapse and vanishing shift, i.e.~$N=1$ and $N^i=0$.} 
 with time function $t=\phi$ and time-flow vector field
$u^\mu\frac{\pa}{\pa x^\mu}=\frac{\pa}{\pa t}$.
The quantity $\chi$, defined above, can be expressed as $\chi=\nabla^\mu u_\mu$ and
represents the expansion of the geodesic congruence generated by $u^\mu$. In FLRW spacetime, one has $\chi=3H$, where $H$ denotes the Hubble rate. It is for this reason that the term $f(\chi)$ in the action (\ref{ActionPrinciple}) plays an important role in the cosmological applications of the model, since 
for a homogenous and isotropic background
$f(\chi)$ reduces to a function of the Hubble rate only. This is  a crucial property of the model, which allows for a straightforward theory reconstruction procedure, starting from a given cosmological background evolution. This aspect will be analyzed in detail in Section~\ref{Sec:Reconstruction}.

It is worth stressing that, although the action for mimetic gravity includes higher-derivative terms through $f(\chi)$, the equations of motion are second order. In fact, mimetic gravity is a particular case of so-called degenerate higher-order scalar tensor theories (DHOST), which are characterised by the absence of Ostrogradski ghost \cite{Langlois:2018jdg,Langlois:2018dxi}. Nevertheless, compared to general relativity, the mimetic gravity theory described by (\ref{ActionPrinciple}) has an extra propagating scalar degree of freedom if $f_{\chi\chi}\neq0$ \cite{Firouzjahi:2017txv,Chamseddine:2014vna}. Importantly, this is always a source of instabilities in the theory, as discussed in Section~\ref{Insta}.

The field equations read as~\cite{Chamseddine:2016uef}
\be\label{FieldEquations}
G_{\mu\nu}=T_{\mu\nu}^{\psi}+\tilde{T}_{\mu\nu} ~,
\ee
where the matter stress-energy tensor is defined as usual
\be
T_{\mu\nu}^{\psi}=\frac{2}{\sqrt{-g}}\frac{\delta S_{\rm m}}{\delta  g^{\mu\nu}} ~,
\ee
and the extra term in Ref.~(\ref{FieldEquations}) is an effective stress-energy tensor arising from the $\phi$-sector of the action (\ref{ActionPrinciple})
\be\label{EffectiveStressEnergy}
\tilde{T}_{\mu\nu}=2\lambda \pa_\mu\phi\pa_\nu\phi+g_{\mu\nu}(\chi f_{\chi}-f+g^{\rho\sigma}\pa_{\rho}f_{\chi}\pa_\sigma\phi)-(\pa_\mu f_{\chi}\pa_\nu\phi+\pa_\nu f_{\chi}\pa_\mu\phi) ~.
\ee
The Lagrange multiplier $\lambda$ can be eliminated by solving the following equation
\be\label{MultiplierEquation}
\Box f_\chi-2\nabla^{\mu}(\lambda\pa_{\mu}\phi)=0 ~,
\ee
which can be obtained by varying the action with respect to $\phi$. Equation~\ref{MultiplierEquation} can be interpreted as a conservation law for the Noether current associated with the global shift-symmetry of the action~(\ref{ActionPrinciple}), see Refs.~\cite{deCesare:2018cts,Mirzagholi:2014ifa}

Considering a flat FLRW model $\de s^2=\de t^2-a^2(t)\delta_{ij}\de x^i\de x^j$, the field equations (\ref{FieldEquations}) lead to a modification of the Friedmann and Raychaudhuri equations
\begin{align}
&\frac{1}{3}\chi^2=\rho+\tilde{\rho}+M ~, \label{FirstFriedmann}\\
&\dot{\chi}=-\frac{3}{2}\left[(\rho+P)+(\tilde{\rho}+\tilde{P})+M \right] ~.\label{SecondFriedmann}
\end{align}
The quantities introduced in Eqs.~(\ref{FirstFriedmann}), (\ref{SecondFriedmann}) are defined as follows: $\rho$ and $P$ denote the energy density and pressure of ordinary matter, whereas $\tilde{\rho}$ and $\tilde{P}$ represent the corresponding quantities for the effective fluid, given by
\begin{align}
\tilde{\rho}&=\chi f_{\chi}-f ~, \label{EffectiveEnergyDensity}\\
\tilde{P}&=-(\tilde{\rho}+ f_{\chi\chi}\dot{\chi}) ~.  \label{EffectivePressure}
\end{align}
The properties of the effective fluid for a quadratic $f(\chi)$ were studied in Ref.~\cite{Mirzagholi:2014ifa}.
Lastly, we have $M=\frac{C}{a^3}$, where $C$ is an integration constant for Eq.~(\ref{MultiplierEquation}). The quantity $M$ represents the energy density of so-called mimetic dark matter \cite{Chamseddine:2013kea}. We note that for vanishing $f$ the action (\ref{ActionPrinciple}) describes irrotational dust minimally coupled to gravity, corresponding to a particular case of the Brown-Kucha\v{r} action \cite{Brown:1994py}.\footnote{See also Refs.~\cite{Husain:2011tk,Husain:2011tm}.}
Finally, we observe that the effective fluid satisfies the continuity equation
\be
\dot{\tilde{\rho}}+\chi(\tilde{\rho}+\tilde{P})=0 ~.
\ee

\section{Theory reconstruction}\label{Sec:Reconstruction}
We henceforth consider a spatially flat, homogeneous and isotropic universe, as described by the FLRW line element $\de s^2=\de t^2-a^2(t)\delta_{ij}\de x^i\de x^j$. The proper time gauge $N=1$ will be used throughout.
The spacetime geometry is then fully characterised by the evolution of a single degree of freedom: the scale factor $a(t)$. Given a theory of gravity with second order field equations, cosmological solutions can be represented as trajectories in the plane~$(a,\chi)$. In general relativity, the trajectories are determined by the Friedmann equation
\be\label{Friedmann}
\frac{1}{3 }\chi^2=\sum_{i}\rho_i ~.
\ee
Here the quantities $\rho_i$ denote the energy density of different matter species. For the sake of simplicity, we can assume that all matter species are non-interacting and have constant equation of state parameters~$w_i$. Thus, we have $\rho_i=c_iV^{-(w_i+1)}$, where $c_i$ are constants depending on the initial conditions and $V=a^3$ is the proper volume of a unit comoving cell.  
It is convenient to introduce a new variable $\eta=V^{-1}$, so that the Friedmann equation can be re-expressed as
\be\label{FriedmannChiEta}
\frac{1}{3 }\chi^2=\sum_{i}c_i\,\eta^{w_i+1} ~.
\ee
Such a parametrization is particularly useful in bouncing cosmologies, where $\eta$ has a bounded range. In the following, we will denote by $\Gamma$ the trajectory in the $(\eta,\chi)$ plane given by Eq.~(\ref{FriedmannChiEta}).

In spite of the derivation given above, based on the standard Friedmann equation, equation~(\ref{FriedmannChiEta})
has a broader applicability. In fact, it also holds in a more general class of modified gravity theories and quantum cosmological models, provided that the corrections to the standard Friedmann equation can be described ---at an effective level--- as perfect fluids. Such effective fluids may have exotic properties and, depending on the model, can violate the energy conditions. This is the case, for instance, in the effective dynamics of both loop quantum cosmology and group field theory condensate cosmology. In fact, having a bounce requires that both the weak and the null energy conditions must be violated due to the effective fluids. The former violation is necessary in order to accommodate for a vanishing expansion, see Eq.~(\ref{FriedmannChiEta}). The latter violation follows instead from the requirement that $\dot{\chi}>0$ at the bounce, and from the Raychaudhuri equation including effective fluids contributions
\be
\dot{\chi}=-\frac{3}{2}\sum_i (\rho_i+P_i) ~.
\ee

\begin{figure}
    \begin{center}
    \subfloat{{\includegraphics[width=.45\columnwidth]{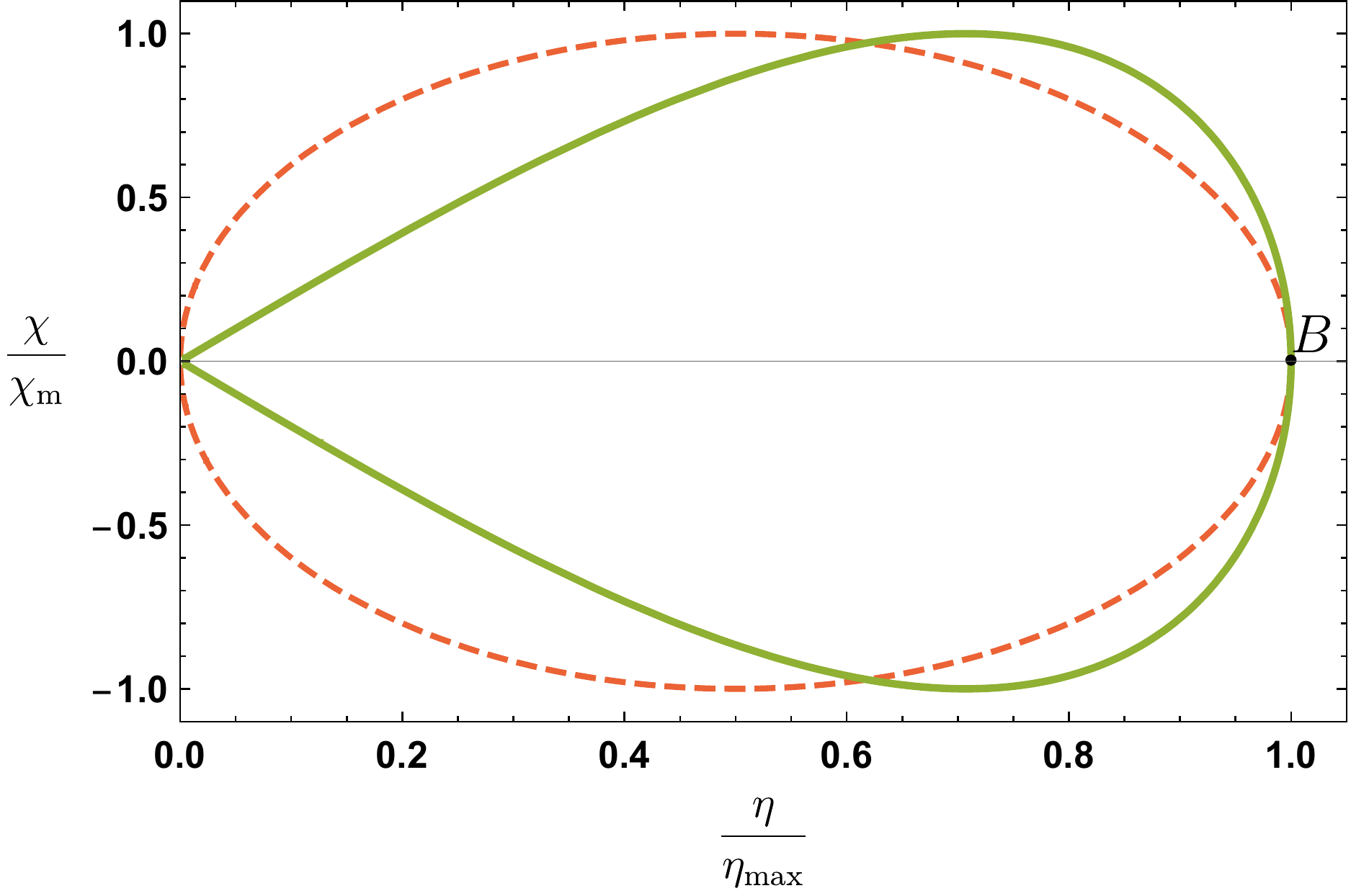} }}
    \qquad
    \subfloat{{\includegraphics[width=.45\columnwidth]{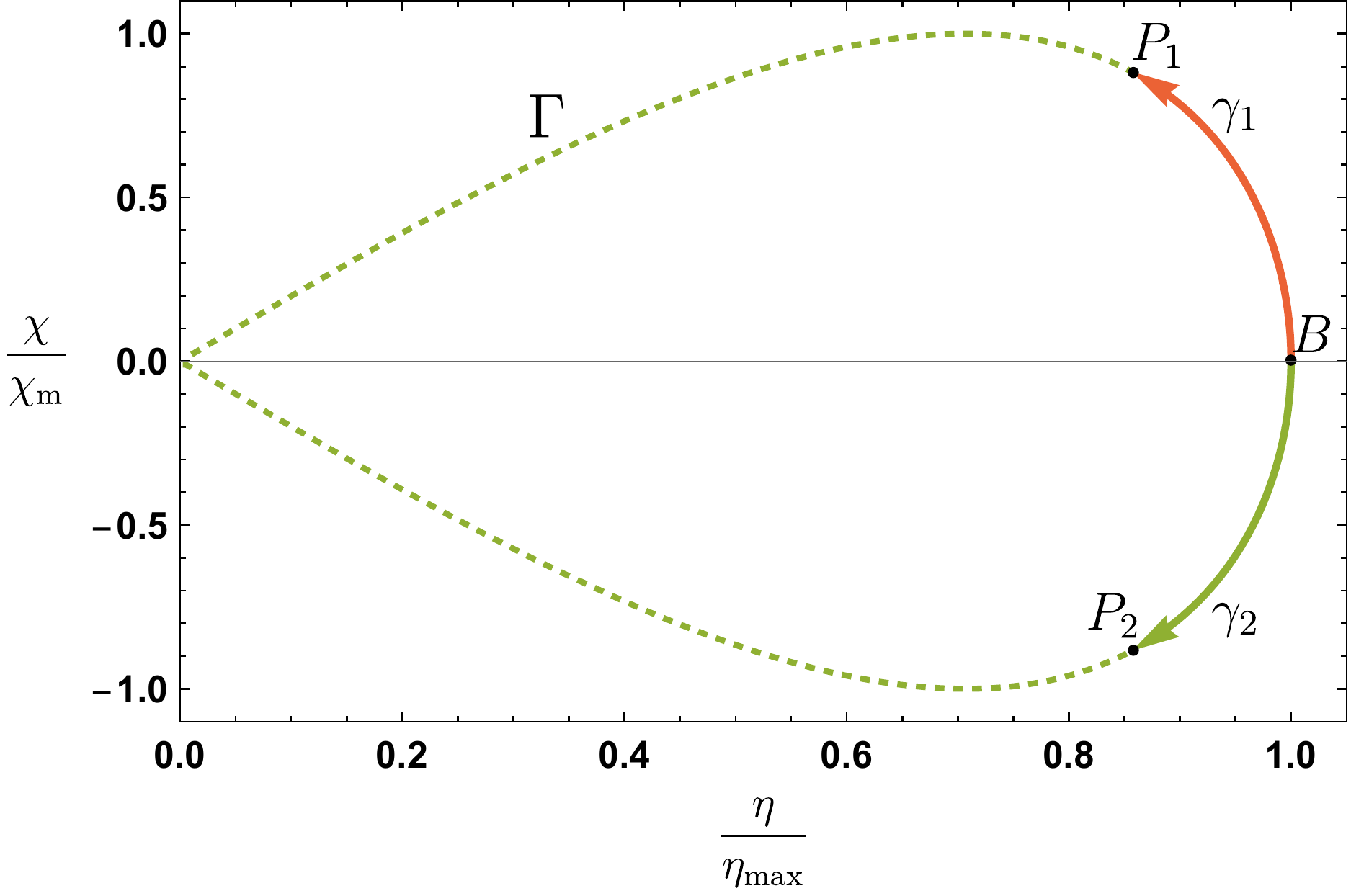} }}
    \caption{Trajectories $\Gamma$ in the $(\eta,\chi)$ plane for a (symmetric) bouncing universe.  The upper half-plane corresponds to the expanding phase, whereas the lower half-plane describes the contracting phase. The bounce is represented by the point $B$, where the expansion rate vanishes and the scale factor attains its minimum (correspondingly $\eta$ is maximised). The left figure shows the trajectory $\Gamma$ for a universe filled with a scalar field (thick green line), or dust (dashed orange line); parameters are chosen so that the two trajectories are characterised by the same critical density $\rho_c$ and limiting expansion rate $\chi_{\rm m}$. The right figure shows the two integration contours $\gamma_1$, $\gamma_2$ used in Eq.~(\ref{OddInt}).}
    \label{FigGamma}
     \end{center}
\end{figure}

It is important to observe that, in general, Equation~(\ref{FriedmannChiEta}) allows to define $\chi$ as a function of $\eta$ only locally. In fact, in bouncing models, the function $\chi(\eta)$ has (at least) two branches. More branches are possible if one allows e.g.~for intermediate recollapse eras; we shall disregard this possibility in the following for simplicity. For a universe undergoing a single bounce, the trajectory $\Gamma$ has the profile depicted in Fig.~\ref{FigGamma}. The bounce is represented by the point $B=(\eta_{\rm max},0)$, where $\Gamma$ and the $\eta$ axis intersect orthogonally. Since we are assuming a flat spatial geometry, both endpoints of $\Gamma$ will have $\eta=0$ if the weak energy condition is satisfied for a large universe. The value of $\chi$ at the endpoints is determined by the equation of state of the dominant matter species in such a regime: for $w>-1$ one has that $\chi$ vanishes as $\eta$ tends to zero, for $w=-1$ (cosmological constant) $\chi$ approaches a constant value. 
We note that for $w+1>0$ the two endpoints coincide with the origin; moreover, for $-1<w<1$ the trajectory $\Gamma$ intersects the $\eta$ axis orthogonally at the origin, whereas for $w\geq1$ it has a cusp.

\subsection{Recostruction procedure}
Given a background evolution as specified by the trajectory $\Gamma$, it is possible to apply a reconstruction procedure that allows to uniquely determine the function $f(\chi)$ in the mimetic gravity action~(\ref{ActionPrinciple}). The method illustrated in this section extends to a generic background evolution the procedure applied in Refs.~\cite{deCesare:2018cts,deHaro:2018sqw,deHaro:2018cpl} and ensures that appropriate matching conditions are implemented at the branching points.\footnote{We note that a different version of mimetic gravity is considered in Ref.~\cite{deHaro:2018cpl} that agrees at the background level with the one presently considered. However, the two theories will differ in general at the level of perturbations.}
We start by rewriting Eq.~(\ref{FirstFriedmann}), using Eq.~(\ref{EffectiveEnergyDensity}), as
\be
\frac{\chi^2}{3}\left[1-3\frac{\de}{\de\chi}\left(\frac{f}{\chi}\right)\right]=\rho ~.
\ee
The solution to this equation can be obtained by quadrature, and is given by
\be\label{SolutionGeneral}
f(\chi)=\frac{\chi}{3} \underset{\gamma}{\int_A^P}\de\chi\;\left(1-\frac{3\rho(\eta(\chi))}{\chi^2}\right)+\bar{c}\,\chi ~,
\ee
where $\bar{c}$ is an integration constant. The integral is computed along an arc of curve $\gamma\subseteq\Gamma$ with endpoints $A$ and $P$, representing a fixed reference point and a generic point on $\Gamma$, respectively. 

In bouncing cosmologies, the background dynamics is characterised by the existence of a limiting curvature scale, which is attained at the bounce. In turn, this scale is related to the existence of a maximum expansion rate, which will be denoted by $\chi_{\rm m}\equiv\underset{\Gamma}{\max}\,\chi$, see Fig.~\ref{FigGamma}. In this class of models, it is convenient to take the bounce as a reference point, i.e.~$A\equiv B$ in Eq.~(\ref{SolutionGeneral}).
Since the energy density of matter is given as a function of the inverse volume, i.e.~$\rho=\rho(\eta)$, the explicit computation of the integral (\ref{SolutionGeneral}) requires the determination of the inverse function $\eta(\chi)$. In general, such an inverse function exists only locally. This implies that in bouncing models the function $f(\chi)$ given by Eq.~(\ref{SolutionGeneral}) must be multivalued as a function of $\chi$.\footnote{We observe that for models entailing a single bounce, the solution (\ref{SolutionGeneral}) is single-valued if regarded as a function of the pair $(\chi,\dot{\chi})$.}
More precisely, in models with a single bounce $f(\chi)$ has two branching points where $\chi$ attains its extrema, one in the expanding phase, the other in the contracting phase. For a generic bouncing background $f(\chi)$ would have three branches, each corresponding to one of the three branches of the inverse functions $\eta(\chi)$. Thus, one branch $f_{\rm\scriptscriptstyle B}$ corresponds to the bounce phase, and two (a priori distinct) branches $f_{\rm\scriptscriptstyle L}^c$, $f_{\rm\scriptscriptstyle L}^e$ correspond to the regions away from the bounce in the contracting and expanding phase, respectively. We will refer to the latter as the large universe branches, characterised by  $\dot{\chi}<0$.
As shown in Section~\ref{Sub:Bounce}, for symmetric bounces the two branches $f_{\rm\scriptscriptstyle L}^c$, $f_{\rm\scriptscriptstyle L}^e$ can be identified, provided that an appropriate choice is made for the integration constant in Eq.~(\ref{SolutionGeneral}).
 
 We remark that our solution for $f$ is continuous on $\Gamma$ by construction. The derivative $f_\chi$ is also continuous, except at the origin $\chi=\eta=0$.\footnote{The fact that the origin is a singular point in the parametrization adopted here should not be too surprising: in fact, it corresponds to the infinite volume limit of both contracting and expanding branches. In a flat universe these are clearly two disconnected regimes.} This ensures that the energy density of the effective fluid, Eq.~(\ref{EffectiveEnergyDensity}), is continuous throughout cosmic history. Thus, the matching conditions prescribed in Ref.~\cite{deCesare:2018cts} are automatically implemented in Eq.~(\ref{SolutionGeneral}).
  As a general property of this class of models $f_{\chi\chi}$ diverges at the branching points, see discussion in Section~\ref{EffectiveGFT}.

After computing the integral in Eq.~(\ref{SolutionGeneral}), the reconstructed action for mimetic gravity can then be obtained by replacing $\chi\to\Box\phi$ in the result. Clearly, the value of the integration constant $\bar{c}$ has no influence on the equations of motion, since the linear term contributes a total divergence to the action~(\ref{ActionPrinciple}). 

\subsection{Bounce asymptotics}\label{Sub:Bounce}
For a symmetric bounce model, the function $f(\chi)$ is even, provided that an appropriate choice of the integration constant is made in Eq.~(\ref{SolutionGeneral}). In fact, defining $P_1=(\eta,\chi)$ and $P_2=(\eta,-\chi)$, with $\eta$ and $\chi$ satisfying the background equation, one has 
\be\label{OddInt}
\underset{\gamma_1}{\int_B^{P_1}}\de\chi\;\left(1-\frac{3\rho(\eta(\chi))}{\chi^2}\right)=-\underset{\gamma_2}{\int_B^{P_2}}\de\chi\;\left(1-\frac{3\rho(\eta(\chi))}{\chi^2}\right)~.
\ee
Thus, the integral is odd. 
The curves $\gamma_1$ and $\gamma_2$ are depicted in Fig.~\ref{FigGamma}. Using Eqs.~(\ref{OddInt}),~(\ref{SolutionGeneral}), it is then straightforward to show that setting $\bar{c}=0$ leads to $f(\chi)=f(-\chi)$. In the following, we shall restrict our attention to symmetric bounce models and assume that $f(\chi)$ be even, unless otherwise stated.

The value of the function $f$ at the bounce is independent from all other details of cosmic history. It can be computed as a limit of Eq.~(\ref{SolutionGeneral}). Denoting by $f_{\rm\scriptscriptstyle B}$ the bounce branch of the multivalued function $f$, we have
\be\label{ValueAtB}
f_{\rm\scriptscriptstyle B}(0)=\lim_{P\to B} \frac{\chi}{3} \underset{\gamma}{\int_B^P}\de\chi\;\left(1-\frac{3\rho(\eta(\chi))}{\chi^2}\right)=\lim_{\chi\to0} \frac{\chi}{3} \int_0^\chi\de\chi\;\left(1-\frac{3\rho(\eta(\chi))}{\chi^2}\right)=\rho_{\rm c} ~,
\ee
where $\rho_{\rm c}$ is the critical density, i.e.~the maximum of the energy density of matter, which is attained at the bounce. Since $f_{\rm\scriptscriptstyle B}$ is even by hypothesis, we have for $\chi\simeq0$
\be\label{GeneralFuncExpAtB}
f_{\rm\scriptscriptstyle B}(\chi)=\rho_{\rm c}+\frac{1}{2!}\vartheta\;\chi^2+\mathcal{O}(\chi^4) ~,
\ee
where we introduced the notation $\vartheta=(f_{\rm\scriptscriptstyle B})_{\chi\chi}\big|_0$~. Hence, it follows that the energy density of the effective fluid at the bounce is given by $\tilde{\rho}=-\rho_{\rm c}$. 
The sign of the second derivative can be determined by the requirement that the effective fluid must violate also the null energy condition (NEC) at the bounce. In fact, using Eq.~(\ref{EffectivePressure}) we have
\be
\tilde{\rho}+\tilde{P}=-(f_{\rm\scriptscriptstyle B})_{\chi\chi}\;\dot{\chi}<0
\ee
Since $\dot{\chi}>0$ at the bounce, we conclude $\vartheta>0$.

NEC violation also allows to derive an upper bound for the duration of the bounce in limiting curvature mimetic gravity. In order to prove such a statement, let us assume that at the bounce the most relevant contributions to  the energy density are due to the effective fluid and to a perfect fluid with equation of state parameter $w$. The condition $\dot{\chi}>0$, which must be valid in a neighbourhood of the bounce, implies
\be\label{NECviolation}
\rho+p+\tilde{\rho}+\tilde{P}<0 ~.
\ee
In turn, Eq.~(\ref{NECviolation}) implies
\be\label{NECviolation_contd}
(1+w)\rho_{c}-(f_{\rm\scriptscriptstyle B})_{\chi\chi}\dot{\chi}<0 ~.
\ee
For first-order bounce models\footnote{The order of the bounce is defined as the positive integer $n$ such that $a^{(2n)}(t_{\rm\scriptscriptstyle B})>0$ is the lowest-order non-vanishing derivative of the scale factor at the bounce \cite{Cattoen:2005dx}.} during the bounce phase the expansion $\chi$ is well approximated by a linear function of time. We can estimate the time derivative of $\chi$ at the bounce as $\dot{\chi}\sim \frac{\chi_{\rm m}}{T}$, where $T$ is the bounce duration. Therefore, in this case we obtain from Eq.~(\ref{NECviolation_contd})
\be\label{BounceTimeScale}
T\lesssim\frac{\vartheta\;\chi_{\rm m}}{\rho_c (1+w)} ~.
\ee
Typically $\rho_c\sim \chi_{\rm m}^2$ and $\vartheta\sim\mathcal{O}(1)$, so that $T\lesssim \chi_{\rm m}^{-1}$. When such an approximation applies, the number of e-folds of expansion during the bounce phase is $N=\log\left(\frac{a(T)}{a_{\rm\scriptscriptstyle B}}\right)\lesssim\mathcal{O}(1)$.
These considerations also apply to the models studied in Section~\ref{EffectiveGFT} (see Eq.~\ref{TaylorBounce} for the corresponding expansion of $f$ near the bounce). In fact, the estimate (\ref{BounceTimeScale}) is in agreement with the upper bound for the number of e-folds obtained in Ref.~\cite{deCesare:2016rsf} for the so-called non-interacting model. We mention that the so-called fast-bounce models, considered e.g.~in Ref.~\cite{Lin:2010pf}, are first-order bounces whose duration is much shorter than the time-scale linked to the maximum expansion rate, i.e.~such that~$T\ll\chi_{\rm m}^{-1}$; such a scenario can be realised in mimetic gravity by requiring $\frac{(f_{\rm\scriptscriptstyle B})_{\chi\chi}}{f_{\rm\scriptscriptstyle B}}\Big|_{\chi=0}\ll \chi_{\rm m}^{-2}$.

\subsection{Late time asymptotics}
Considerations on the evolution of the universe at late times 
allow to put restrictions
on the leading order terms of the branch $f_{\rm\scriptscriptstyle L}$ around $\chi\simeq0$.
In fact, we observe that the effective fluid is characterised by a time-dependent equation of state parameter $\tilde{w}$, given by 
\be
\tilde{w}=\frac{\tilde{P}}{\tilde{\rho}}=-\left(1+\frac{f_{\chi\chi}}{\tilde{\rho}}\dot{\chi}\right)~,
\ee
where we used Eqs.~(\ref{EffectiveEnergyDensity}), (\ref{EffectivePressure}). It is interesting to examine the case where the universe at late times is dominated by matter with equation of state $w$ and the effective fluid is sub-dominant, with $\tilde{w}$ approaching a constant value as $\chi\to0$. Clearly, consistency of such assumptions requires $w<\tilde{w}$. The leading order term in the expansion of $f_{\rm\scriptscriptstyle L}(\chi)$ around $\chi\simeq 0$ is then given by
\be\label{LateTimesAsympt}
f_{\rm\scriptscriptstyle L}(\chi)\simeq\lambda\; \chi^{2\left(\frac{\tilde{w}+1}{w+1}\right)}~,
\ee
where $\lambda$ is a constant. In fact, since by hypothesis we have to leading order $\chi\sim\eta^{\frac{1+w}{2}}$,
Eq.~(\ref{LateTimesAsympt}) implies $\tilde{\rho}\sim \eta^{1+\tilde{w}}$, consistently with our assumptions.

\section{Effective approach to quantum gravitational bouncing cosmologies}\label{EffectiveGFT}
In Ref.~\cite{deCesare:2018cts} the reconstruction procedure outlined in Section~\ref{Sec:Reconstruction} was successfully applied to the cosmological dynamics obtained from group field theory condensates in \cite{\GFTbounce}. The evolution equation for such a model can be expressed in relational form by introducing a minimally coupled massless scalar field $\psi$ \cite{Gielen:2018fqv}. In fact, provided that its momentum be non-vanishing $p_{\psi}\neq0$, $\psi$ is a monotonic function of $t$ and thus represents a perfect clock. For definiteness, we will assume $p_{\psi}>0$. Using the relational clock $\psi$ as time, the FLRW line element can be expressed as
\be
\de s^2=N^2(\psi)\, \de\psi^2-a^2(\psi) \delta_{ij}\de x^i\de x^j ~,
\ee
where the lapse function reads as
\be
N(\psi)=(\dot{\psi})^{-1}=p_\psi^{-1}\,a^3(\psi) ~.
\ee
We can define a relational Hubble rate as $\mathcal{H}=\frac{a^{\prime}}{a}$,
where a prime denotes differentiation with respect to $\psi$. The expansion $\chi$ is related to $\mathcal{H}$ as follows
\be\label{Eq:RelationChi_H}
\chi=3\,p_\psi \frac{\mathcal{H}}{a^3}~.
\ee

The relational Friedmann equation governing the dynamics of group field theory condensates reads as (recall $V=a^3$)
\be\label{RelationalFriedmann}
\mathcal{H}^2=\frac{1}{6}+\frac{\alpha}{V}-\frac{\beta}{V^2} ~,
\ee
where $\alpha$ and $\beta>0$ are parameters depending on the details of the microscopic model, see Ref.~\cite{\GFTbounce}.\footnote{It is worth remarking that $\alpha$ and $\beta$ are defined only up to arbitrary constant rescalings of the comoving volume $V_0$, which was set equal to one above. We have in general $V=V_0\, a^3$. Under the transformation $V_0\to k V_0$ with constant $k$, $\alpha$ and $\beta$ transform according to $\alpha\to k\,\alpha$, $\beta\to k^2 \beta$. Thus, the scale invariance property of the standard Friedmann equation is preserved by the quantum corrections. In the group field theory formalism such rescaling properties correspond to the invariance of the dynamics under constant rescalings of the number of quanta, cf.~Ref.~\cite{\GFTbounce}.}
An effective Friedmann equation with the same form as Eq.~(\ref{RelationalFriedmann}) was obtained in the group field theory models of Refs.~\cite{Adjei:2017bfm,Wilson-Ewing:2018mrp}.
 The first term in Eq.~(\ref{RelationalFriedmann}) is the contribution of the massless scalar field $\psi$, whereas the remaining two terms represent quantum gravitational corrections; in particular, the $\alpha$ term represents a correction to the effective dynamics of loop quantum cosmology. It must be stressed that, for simplicity, we are neglecting interactions between group field theory quanta, which would contribute additional terms to Eq.~(\ref{RelationalFriedmann}). The cosmological consequences of interactions were considered in Ref.~\cite{deCesare:2016rsf}.

Changing time parametrization back to proper time and recalling $\eta=V^{-1}$, we have
\be\label{ChiOfV}
\frac{1}{3}\chi^2=p_\psi^2\left(\frac{1}{2}\,\eta^2+3\alpha\, \eta^3-3\beta\,\eta^4\right) ~.
\ee
The first term to the r.h.s.~of Eq.~(\ref{ChiOfV}) gives the energy density $\rho_\psi$ of the scalar $\psi$; the quantum gravitational corrections (second and third terms) correspond instead to effective fluids with equation of state parameter $w=2\,,\,3$. The third term becomes important for large values of $\eta$ (i.e.~small values of the scale factor); moreover, since $\beta>0$ such a term violates both the weak and the null energy conditions, and is therefore responsible for the bounce. It must be noted that the bounce is symmetric for any choice of parameters in this model.
The equation for $\dot{\chi}$ is
\be
\dot{\chi}=-\frac{3}{2}p_\psi^2\left(\eta^2 +9\alpha\,\eta^3-12\beta\,\eta^4 \right) ~.
\ee
For further details on the effective fluid description of quantum gravity corrections in the effective Friedmann equation arising in the group field theory approach, including interactions between quanta, the reader is referred to Refs.~\cite{deCesare:2016axk,deCesare:2016rsf}. For a large universe (i.e.~small $\eta$) the first term in Eq.~(\ref{ChiOfV}) becomes the dominant one: the standard Friedmann evolution is thus recovered, and the quantum gravity corrections are sub-leading.

The background evolution (\ref{ChiOfV}) can be exactly reproduced in mimetic gravity if the function $f(\chi)$ is given by \cite{deCesare:2018cts}
\be\label{SolutionMimeticFunction}
 f(\chi)=\rho_\psi(\chi)+\frac{1}{3}\chi^2+\frac{p_\psi}{3\sqrt{\beta}}|\chi|\left[ \arctan \left(\frac{1}{\sqrt{\beta}}\frac{\de |\mathcal{H}|}{\de \eta} \right)   +\frac{\pi}{2}\right] ~.
\ee
By construction, the different branches of the multivalued function in Eq.~(\ref{SolutionMimeticFunction}) satisfy matching conditions at the branching points, so as to ensure the regularity of cosmological evolution. Around the bounce the following expansion holds
\be\label{TaylorBounce}
f_{\scriptscriptstyle\rm B}(\chi)=\rho_c+\frac{1}{3}\left(\frac{2 V_{\scriptscriptstyle\rm B}+3\alpha}{V_{\scriptscriptstyle\rm B}+3\alpha}\right)\chi^2+\mathcal{O}(\chi^4) ~,
\ee
where $V_{\scriptscriptstyle\rm B}=-3\alpha+\sqrt{9\alpha^2+6\beta}$ is the volume at the bounce and $\rho_c=\frac{p_\psi^2}{2V_{\sscr\rm B}^2}$. For the asymptotic expansion of $f(\chi)$ around the branching points at maximum expansion rate $|\chi|=\chi_{\rm m}$, see Ref.~\cite{deCesare:2018cts}. Both $f$ and $f_{\chi}$ are continuously matched at the branching points. However, the second derivative $f_{\chi\chi}$ has a discontinuity there: this is a general property of mimetic gravity theories with a limiting curvature scale. Nevertheless, the effective pressure $\tilde{P}(\chi)$ is guaranteed to be finite even when $f_{\chi\chi}$ diverges, 
since Eq.~(\ref{SecondFriedmann}) implies
\be
\tilde{P}(\pm\chi_{\rm m})=-(\rho+P)=-(w+1)\rho~.
\ee

 When the universe is large (i.e.~in the regime $\chi\,,\,\eta\sim0$) one has the expansion (disregarding the linear term, which does not affect the equations of motion)
\be
f_{\scriptscriptstyle\rm L}(\chi)=\sqrt{\frac{2}{3}}\frac{\alpha}{p_\psi}|\chi|^3-\frac{4}{p_\psi^2}\left(\alpha^2+\frac{1}{9}\beta\right)\chi^4+\mathcal{O}(|\chi|^5) ~,
\ee
which can be rewritten as
\be\label{TaylorLate}
f_{\scriptscriptstyle\rm L}(\chi)=\frac{\alpha}{2\, V_{*}}\sqrt{2+\frac{6\alpha}{V_{*}}}\,\frac{|\chi|^3}{\chi_{\rm m}}-\frac{(V_{*}+3\alpha)(V_{*}^2+9\alpha V_{*}+108\alpha^2)}{36V_{*}^3}\frac{\chi^4}{\chi_{\rm m}^2}+\mathcal{O}(|\chi|^5) ~,
\ee
where $\chi_{\rm m}=\frac{p_\psi}{2V_*}\sqrt{3+\frac{9\alpha}{V_{*}}}$, and $V_{*}=\frac{1}{2}\left(\sqrt{81\alpha^2+48\beta}-9\alpha\right)$ is the volume at $\chi=\chi_{\rm m}$. Note that, if $\alpha=0$, the next non-vanishing term in the expansion is $\mathcal{O}(\chi^6)$.

Once the function $f(\chi)$ has been reconstructed from a given background evolution, one can also consider different matter species coupled to gravity. It must be pointed out that, when matter species other than a minimally coupled massless scalar field are considered, parameters such as $p_\psi$ and $V_{\scriptscriptstyle\rm B}$ in Eq.~(\ref{SolutionMimeticFunction}) 
lose their usual interpretation. This is to be expected, since the relation between $\chi$ and $\eta$ will be different from Eq.~(\ref{ChiOfV}) in the general case. Nevertheless, the values of the critical energy density $\rho_c$ and the maximum expansion rate $\chi_{\rm m}$ are not affected by the different matter species, and represent universal features of the model. 

Let us now assume hydrodynamic matter with constant equation of state parameter $w$. Comparing Eqs.~(\ref{TaylorLate}) and (\ref{LateTimesAsympt}), at late times we obtain a simple description of the effective fluid corresponding to the mimetic gravity corrections as a sum of perfect fluid contributions, each with a constant equation of state. Specifically, we find for the third order term in Eq.~(\ref{TaylorLate}) $\tilde{w}_3=\frac{1}{2}(3w+1)$, whereas for the fourth order term we have $\tilde{w}_4=2w+1$. Clearly, for a massless scalar field $w=1$ one recovers the 
effective fluid corrections given in Eq.~(\ref{ChiOfV}). 

\subsection{A special case: reproducing the LQC effective dynamics}
The case $\alpha=0$ is special and deserves being discussed separately. In fact, in this case one recovers the model of Ref.~\cite{Chamseddine:2016uef}, which reproduces the effective dynamics of loop quantum cosmology for a spatially flat, isotropic universe. After locally inverting $\chi=\chi(\eta)$, one finds the two branches of the function $f(\chi)$
\begin{align}
f_{\rm\scriptscriptstyle B}&=\frac{2}{3}\chi_{\rm m}^2\left\{1+\frac{1}{2}q^2+\sqrt{1-q^2 } +q\arcsin(q)     \right\}  ~, \label{Eq:LQCbounceBranch} \\
f_{\rm\scriptscriptstyle L}&=\frac{2}{3}\chi_{\rm m}^2\left\{1+\frac{1}{2}q^2-\sqrt{1-q^2}  -|q|\big(\arcsin|q| -\pi \big)     \right\} ~, \label{Eq:LQClateBranch}
\end{align}
where $\chi_{\rm m}=p_\psi\sqrt{\frac{3}{48\beta}}$ and  we defined $q=\frac{\chi}{\chi_{\rm m}}$ to make the notation lighter. It must be noted that Eqs.~(\ref{Eq:LQCbounceBranch}), (\ref{Eq:LQClateBranch}) do not make any reference to the scalar field $\psi$, which was assumed as the only matter species coupled to gravity in the derivation of Eq.~(\ref{RelationalFriedmann}) in Ref.~\cite{\GFTbounce}. Thus, for $\alpha=0$ the effective Friedmann equation will take the same universal form regardless of the matter species considered. Using Eq.~(\ref{ValueAtB}), the critical energy density is determined as $\rho_c=f_{\rm\scriptscriptstyle B}(0)=\frac{4}{3}\chi_{\rm m}^2$. The energy density of the effective fluid can be computed using Eq.~(\ref{EffectiveEnergyDensity}); the result is $\tilde{\rho}=-\frac{\rho_c}{2}\left(1-\frac{q^2}{2}\pm\sqrt{1-q^2}\right)$, where the upper sign corresponds to the bounce branch and the lower one corresponds to a large universe. After some straightforward algebraic manipulations the Friedmann equation (\ref{FirstFriedmann}) can then be recast in the following form
\be\label{EffectiveLQC}
\frac{1}{3}\chi^2=\rho\left(1-\frac{\rho}{\rho_c}\right) ~,
\ee
where $\rho$ denotes the total energy density of all matter species that are present. Similarly, using Eqs.~(\ref{SecondFriedmann}) and (\ref{EffectivePressure}) we can obtain the equation for $\dot{\chi}$. We have, for a general $f(\chi)$
\be\label{DotChiEqinter}
\left(1-\frac{3}{2}f_{\chi\chi}\right)\dot{\chi}=-\frac{3}{2}(\rho+P) ~.
\ee
The bracket to the r.h.s.~of Eq.~(\ref{DotChiEqinter}) can be evaluated using Eqs.~(\ref{Eq:LQCbounceBranch}), (\ref{Eq:LQClateBranch}) 
\be\label{DenSoundSpeedLQC}
1-\frac{3}{2}f_{\chi\chi}=\mp \frac{1}{\sqrt{1-q^2}}=\left(1-\frac{2\rho}{\rho_c}\right)^{-1} ~,
\ee
where we used Eq.~(\ref{EffectiveLQC}) in the last equality. Finally, we have
\be\label{DotChiLQC}
\dot{\chi}=-\frac{3}{2}(\rho+P)\left(1-\frac{2\rho}{\rho_c}\right) ~.
\ee
Thus, the time derivative of the expansion is positive for $\frac{\rho_c}{2}<\rho\leq \rho_c$ (super-inflation). This is to be contrasted with general relativity, where one always has $\dot{\chi}<0$ for matter satisfying the null energy condition.
Equations~(\ref{EffectiveLQC}) and (\ref{DotChiLQC}) coincide with the effective dynamics of (flat, isotropic) loop quantum cosmology, see e.g.~Ref.~\cite{Ashtekar:2011ni}.

It is important to observe that one must change branch of $f(\chi)$ when $\dot{\chi}=0$ \cite{Brahma:2018dwx}. This happens when the density reaches the value $\frac{\rho_c}{2}$, see Eq.~(\ref{DotChiLQC}), whereby the expansion attains its extremum $\chi^2=\chi_{\rm m}^2$. It must be noted that in both branches, as given by Eqs.~(\ref{Eq:LQCbounceBranch}),~(\ref{Eq:LQClateBranch}), $f_{\chi\chi}$ diverges as $|\chi|\to\chi_{\rm m}$; however, the effective pressure $\tilde{P}$ is continuous in the limit since $\tilde{P}=-\frac{\rho}{\rho_c}\left(\rho+2P\right)$.

Exact solutions of the effective Friedmann equation (\ref{EffectiveLQC}) can be derived for hydrodynamic matter (see Ref.~\cite{Chamseddine:2016uef})
\be
a(t)=a_{\sscr\rm B}\left(1+\frac{3}{4}\rho_c(w+1)^2(t-t_{\sscr\rm B})^2 \right)^{\frac{1}{3(1+w)}} ~,
\ee
where the origin of time has been set so as to have the bounce at $t=0$. Provided that matter satisfies the null energy condition, one finds for the bounce duration (defined so as to have $\chi(T)=\chi_{\rm m}$)
\be
T=\frac{1}{\chi_{\rm m}(1+w)} ~,
\ee
which is in good agreement with the estimate given by Eq.~(\ref{BounceTimeScale}).

Finally, the expansions (\ref{TaylorBounce}) and (\ref{TaylorLate}) for $\alpha=0$ become, respectively
\be
f_{\scriptscriptstyle\rm B}(\chi)=\rho_c+\frac{2}{3}\chi^2+\mathcal{O}(\chi^4) ~,
\ee
and
\be\label{LQClateAsymptotics}
f_{\scriptscriptstyle\rm L}(\chi)=-\frac{1}{36}\frac{\chi^4}{\chi_{\rm m}^2}+\mathcal{O}(\chi^6) ~.
\ee

\section{Anisotropies near the bounce}\label{Anisotropies}
In this Section we generalise the analysis of Ref.~\cite{Chamseddine:2016uef}, studying the evolution of anisotropies near the bounce in a non-singular Bianchi~I spacetime, for the model of Section~\ref{EffectiveGFT} and in the presence of hydrodynamic matter with generic equation of state.

The line element of Bianchi~I in proper time gauge is
\be
\de s^2=\de t^2-a^2(t)\sum_i \e^{2\beta_{(i)}(t)}(\de x^i)^2 ~,
\ee
where $a(t)$ is the mean scale factor, and the variables $\beta_{(i)}$ representing the anisotropies satisfy $\sum_{i}\beta_{(i)}=0$. We will assume hydrodynamical matter with barotropic equation of state. Using the field equations~(\ref{FieldEquations}), it can be shown that the $\beta_{(i)}$ evolve according to
\be\label{EqForBeta}
\ddot{\beta}_{(i)}+\chi\, \dot{\beta}_{(i)}=0 ~.
\ee
The solution of Eq.~(\ref{EqForBeta}) gives
\be\label{SolBetaDot}
\dot{\beta}_{(i)}=\frac{\lambda_{(i)}}{a^3(t)} ~,
\ee
with $\lambda_{(i)}$ integration constants satisfying $\sum_{i}\lambda_{(i)}=0$. The field equations lead to an effective Friedmann equation for the mean scale factor, which includes the contribution of anisotropies
\be\label{FriedmannAnisotrop}
\frac{1}{3}\chi^2=\rho+\tilde{\rho}+\frac{1}{2}\sum_{i}\dot{\beta}_{(i)}^2 ~.
\ee
The last term of Eq.~(\ref{FriedmannAnisotrop}) represents the effective energy density of anisotropies (cf.~e.g.~Ref.~\cite{Cai:2013vm}), which will be denoted by $\rho_{\sscr\Sigma}$. Using Eq.~(\ref{SolBetaDot}), we have 
\be
\rho_{\sscr\Sigma}=\frac{\Sigma^2}{2a^6} ~,
\ee
having defined the shear scalar as $\Sigma^2=\sum_{i}\lambda_{(i)}^2$. Thus, the contribution of anisotropies to the modified Friedmann equation is described as a perfect fluid with stiff equation of state $w=1$, as in general relativity.

The evolution of anisotropies, as represented by the $\beta_{(i)}$, is obtained by integrating Eq.~(\ref{SolBetaDot})
\be\label{SolBetaIntegral}
\beta_{(i)}(t)=\lambda_{(i)}\int\frac{\de t}{a^{3}(t)} ~,
\ee
where $a(t)$ in the integrand is a solution of Eq.~(\ref{FriedmannAnisotrop}). In the remainder of this Section, we will determine the evolution of anisotropies during the bounce phase for the function $f(\chi)$ given by Eq.~(\ref{SolutionMimeticFunction}). Since we are only interested in the region around the bounce, it is convenient to use the expansion (\ref{TaylorBounce}). The energy density of the effective fluid then reads as
\be
\tilde{\rho}\simeq-\rho_c+\frac{1}{3}\left(\frac{2 V_{\scriptscriptstyle\rm B}+3\alpha}{V_{\scriptscriptstyle\rm B}+3\alpha}\right)\chi^2 ~.
\ee
The effective Friedmann equations in this regime can then be recast as
\begin{align}
\frac{\chi^2}{3}&\simeq\left(\frac{V_{\sscr\rm B}+3\alpha}{V_{\sscr\rm B}}\right)\left(\rho_c-\rho-\rho_{\sscr\Sigma}\right) ~,\label{FriedmannApproxBounce1} \\
\dot{\chi}&\simeq\frac{3}{2}\left(\frac{V_{\sscr\rm B}+3\alpha}{V_{\sscr\rm B}}\right)\left(\rho+p+2\rho_{\sscr\Sigma}\right) ~.
\end{align}

At the bounce the scale factor attains its minimum $a_{\sscr\rm B}$, and the r.h.s.~of Eq.~(\ref{FriedmannApproxBounce1}) must vanish. We can use this condition to determine the energy density of matter at the bounce $\rho_{\sscr B}$ (not to be confused with the critical energy density $\rho_c$, which includes the contribution of anisotropies). We have
\be\label{SolRhoB}
\rho_{\sscr B}=\rho_c-\rho_{\sscr \Sigma ,\rm  B} ~,
\ee
with $\rho_{\sscr \Sigma ,\rm  B}=\frac{\Sigma^2}{2a_{\sscr\rm B}^6}$ being the energy density of anisotropies at the bounce.
The r.h.s.~of Eq.~(\ref{FriedmannApproxBounce1}) can be expanded around $a_{\sscr\rm B}$; taking into account that $\rho=\rho_{\sscr B}\left(\frac{a_{\sscr B}}{a}\right)^{3(w+1)}$, this gives
\be\label{FriedmannApproxBounce2}
\frac{\chi^2}{3}\simeq3\left(\frac{V_{\sscr\rm B}+3\alpha}{V_{\sscr\rm B}}\right)\left(\rho_{\sscr B} (w+1)+2\rho_{\sscr \Sigma ,\rm  B}\right)\left(\frac{a}{a_{\sscr\rm B}}-1\right) ~.
\ee
Taking into account Eq.~(\ref{SolRhoB}), we can rewrite Eq.~(\ref{FriedmannApproxBounce2}) as
\be
\frac{\chi^2}{3}\simeq3(w+1)\left(\frac{V_{\sscr\rm B}+3\alpha}{V_{\sscr\rm B}}\right)\left(\rho_{c} -\frac{w-1}{w+1}\rho_{\sscr \Sigma ,\rm  B}\right)\left(\frac{a}{a_{\sscr\rm B}}-1\right) ~.
\ee
The solution is
\be\label{SolScaleFactorBounce}
a(t)\simeq a_{\sscr\rm B}\left(1+\frac{1}{4} \Omega^2 t^2\right) ~,
\ee
where we defined
\be
\Omega^2=(w+1)\left(\frac{V_{\sscr\rm B}+3\alpha}{V_{\sscr\rm B}}\right)\left(\rho_{c} -\frac{w-1}{w+1}\rho_{\sscr \Sigma ,\rm  B}\right) ~.
\ee
The solution (\ref{SolScaleFactorBounce}) for the scale factor shows that, regardless of the presence of anisotropies, the model features a first order bounce, according to the definition given in Ref.~\cite{Cattoen:2005dx}. From Eq.~(\ref{SolScaleFactorBounce}), we find that the mean expansion rate evolves as
\be
\chi(t)\simeq\frac{3}{2}\Omega^2\, t~.
\ee
Finally, using Eqs.~(\ref{SolScaleFactorBounce}) and (\ref{SolBetaIntegral}) we find that the $\beta_{(i)}$ evolve linearly during the bounce
\be\label{BetaEvolution}
\beta_{(i)}(t)\simeq\beta_{(i)}^0+\frac{\lambda_{(i)}}{a_{\sscr\rm B}^3}\,t ~,
\ee
where $\beta_{(i)}^0$ are integration constants. Our solution~(\ref{BetaEvolution}) shows that anisotropies stay bounded during the bounce, and can be kept under control by means of a suitable choice of parameters for the model. It is interesting to compare this result with a similar one obtained in Ref.~\cite{Cai:2013vm} for a non-singular bouncing model based on kinetic gravity braiding theories \cite{Deffayet:2010qz}.

\section{Effective Gravitational Constant(s)}\label{GravConst}
The cosmological background equations of mimetic gravity, Eqs.~(\ref{FirstFriedmann}),~(\ref{SecondFriedmann}), can be recast in an alternative form which makes no reference to perfect fluids. The effects introduced by the function $f(\chi)$ in the action (\ref{ActionPrinciple}) are then included in two effective gravitational ``constants'' $G_{\rm F}^{\sscr eff}$ and $G_{\rm R}^{\sscr eff}$, representing respectively the effective coupling of matter to gravity in the Friedmann and the Raychaudhuri equations
\begin{align}
&\frac{1}{3}\chi^2=8\pi\, G_{\rm F}^{\sscr eff}(\chi)\rho ~,\label{EffFriedmann1} \\
&\dot{\chi}=-12\pi\, G_{\rm R}^{\sscr eff}(\chi)(\rho +P) ~.\label{EffFriedmann2}
\end{align}
The effective couplings are functions of the expansion rate, and are defined as
\begin{align}
&8\pi\, G_{\rm F}^{\sscr eff}(\chi)=\left(1-3\frac{\de}{\de\chi}\left(\frac{f}{\chi}\right)\right)^{-1} ~,\\
&8\pi\, G_{\rm R}^{\sscr eff}(\chi)=\left(1-\frac{3}{2}f_{\chi\chi}\right)^{-1} ~.
\end{align}
It is worth remarking that variable gravitational constants arise in this framework despite of the fact that the action~(\ref{ActionPrinciple}) contains no dilaton couplings. In fact, the reformulation provided here hinges on the presence of a function of the expansion rate $f(\chi)$.

From Eqs.~(\ref{EffFriedmann1}), (\ref{EffFriedmann2}), and the continuity equation for matter, we find the following equation relating the change of $G_{\rm F}^{\sscr eff}$ over time to the difference between the two gravitational constants
\be\label{ConsistencyTwoGeff}
\dot{G}_{\rm F}^{\sscr eff}\rho=\chi (G_{\rm F}^{\sscr eff}-G_{\rm R}^{\sscr eff})(\rho+p) ~.
\ee
We observe that $G_{\rm F}^{\sscr eff}=G_{\rm R}^{\sscr eff}$ if and only if $f(\chi)=k_1\, \chi+ \frac{k_2}{2}\, \chi^2$. In this case, the linear term in $\chi$ has no effect, while the quadratic one leads to a finite redefinition of the Newton constant $8\pi\, G_{\rm F}^{\sscr eff}=\left(1-\frac{3}{2} k_2\right)^{-1}$ (see Ref.~\cite{Mirzagholi:2014ifa}); thus, in a large universe we must require $k_2<\frac{2}{3}$ to ensure that the gravitational interaction remains attractive.\footnote{This must be constrasted with the case of bouncing models examined in Sections~\ref{Sub:Bounce} and \ref{EffectiveGFT}, where the coefficient of the quadratic term must satisfy an opposite inequality in order to guarantee that gravity becomes repulsive at the bounce.} In the general case, both $G_{\rm F}^{\sscr eff}$ and $G_{\rm R}^{\sscr eff}$ will evolve with $\chi$. For instance, assuming that in the large universe branch one has $f(\chi)\simeq k\,\chi^p$ with $p>2$ to leading order in $\chi$, leads to 
\begin{align}
&8\pi\, G_{\rm F}^{\sscr eff}(\chi)\simeq 1+3 k(p-1) \chi^{p-2} ~,\\
&8\pi\, G_{\rm R}^{\sscr eff}(\chi)\simeq 1+\frac{3}{2} k\, p(p-1) \chi^{p-2}~.
\end{align}
If we assume that the universe (away from the bounce) is dominated by hydrodynamic matter with equation of state parameter $w$, we have \begin{align}
&8\pi\, G_{\rm F}^{\sscr eff}(t)\simeq 1+3 k(p-1) \left(\frac{2}{(w+1) t}\right)^{p-2} ~,\label{ModelGeff}\\
&8\pi\, G_{\rm R}^{\sscr eff}(t)\simeq 1+\frac{3}{2} k\, p(p-1) \left(\frac{2}{(w+1) t}\right)^{p-2}~.
\end{align}

The reformulation of the cosmological equations of mimetic gravity offered by Eq.~(\ref{EffFriedmann1}),~(\ref{EffFriedmann2}) suggests that the coefficients of the leading order terms in the expansion of the branch $f_{\sscr\rm L}$ can be constrained using observational bounds on the time variation of the gravitational constant. We have from Eq.~(\ref{ModelGeff}), for a small $k$  and retaining only the main contribution (corresponding to the radiation dominated era, $w=\frac{1}{3}$)
\be
\frac{\Delta G^{\sscr eff}_{\rm F}}{G^{\sscr eff}_{\rm F}}= 1-\frac{G^{\sscr eff}_{\rm F}(t_{\sscr\rm BBN})}{G^{\sscr eff}_{\rm F}(t_{0})}\simeq 
3 k(p-1) \left(\frac{3}{2}\right)^{p-2}(t_{\sscr\rm BBN})^{2-p} ~.
\ee
where $t_0$ is the age of the Universe and  $t_{\sscr\rm BBN}$ is the time of nucleosynthesis.
Bounds on the time variation of the gravitational constant $G^{\sscr eff}_{\rm F}$  can be derived from primordial nucleosynthesis: $-0.10<\frac{\Delta G^{\sscr eff}_{\rm F}}{G^{\sscr eff}_{\rm F}}<0.13$~\cite{Uzan:2010pm,Cyburt:2004yc}. For a given $p>2$, such a bound can be translated into a constraint on $k$. However, such a constraint is very weak for bouncing models.
In fact, if the limiting curvature hypothesis is made, dimensional arguments suggest that $k\sim \chi_{\rm m}^{2-p}$. This is in fact the case for the models considered in Section~\ref{EffectiveGFT}, see~Eqs.~(\ref{LQClateAsymptotics}),~(\ref{TaylorLate}). Moreover, typically one has for the limiting value of the expansion rate $\chi_{\rm m}\sim t_{\sscr\rm Pl}^{-1}$, where $t_{\sscr\rm Pl}$ is Planck time. Therefore, the time variation of the gravitational constant is extremely small in such models $\frac{\Delta G^{\sscr eff}_{\rm F}}{G^{\sscr eff}_{\rm F}}\sim \left(\frac{t_{\sscr\rm Pl}}{t_{\sscr\rm BBN}}\right)^{p-2}$.

A more detailed investigation of the phenomenological consequences of the time variation of $G^{\sscr eff}_{\rm F}$ and $G^{\sscr eff}_{\rm R}$ is beyond the scope of the present article and will be left for future work.

\section{Instabilities}\label{Insta}
Our presentation of mimetic gravity would not be complete without a discussion of perturbative instabilities. Instabilities of cosmological perturbations for the mimetic gravity theory with action~(\ref{ActionPrinciple}) have been studied in Refs.~\cite{Firouzjahi:2017txv,Takahashi:2017pje} for a generic $f(\chi)$; for earlier studies focused on the case of a quadratic $f$ see Ref.~\cite{Ijjas:2016pad,Ramazanov:2016xhp}.\footnote{It must be noted that the quadratic case is equivalent with the IR limit of projectable Ho\v{r}ava-Lifshitz gravity \cite{Ramazanov:2016xhp}, see also Ref.~\cite{Capela:2014xta}.} Compared to general relativity, the theory has one extra propagating scalar degree of freedom, 
whose speed of sound is given by
\be\label{SoundSpeed}
c_s^2=\frac{1}{2}\frac{f_{\chi\chi}}{1-\frac{3}{2}f_{\chi\chi}} ~.
\ee
Depending on the sign of the speed of sound, the theory has a ghost instability (for $c_s^2>0$) or a gradient instability (for $c_s^2<0$), see references above. The propagation speed of tensor perturbations is not affected by the term $f(\chi)$ in the action (\ref{ActionPrinciple}).\footnote{The situation is different in other versions of mimetic gravity, see e.g.~\cite{Langlois:2018jdg} for a general analysis based on the DHOST formulation of (extended) mimetic gravity theories.}

In the following we will assume that the analytic properties of the function $f(\chi)$ are such as to accommodate for a bouncing background. Some general conclusions can then be drawn on the profile of the speed of sound as a function of the expansion, based on the results derived in Section~\ref{Sub:Bounce}. In fact, around the bounce $f(\chi)$ must admit the expansion (\ref{GeneralFuncExpAtB}). Moreover, since $\dot{\chi}>0$ in a neighbourhood of the bounce, Eq.~(\ref{DotChiEqinter}) implies that we must have $\vartheta>\frac{2}{3}$, provided that ordinary matter fields satisfy the NEC. Thus, at the bounce we have
\be
c_s^2= \frac{\vartheta}{2-3\vartheta}<0 ~,
\ee
which corresponds to a gradient instability.
The expansion rate attains its extremum at $|\chi|=\chi_{\rm m}$, where two different  branches of the multivalued function $f(\chi)$ are joined together; at that point the second derivative $f_{\chi\chi}$ is divergent, whereby the speed of sound squared takes the universal value $c_s^2= -\frac{1}{3}$. We conclude that a generic feature of bouncing models in mimetic gravity is that the bounce is always accompanied by a gradient instability of scalar perturbations, which extends beyond the onset of the standard decelerated expansion. The possibility that $c_s^2$ may turn to positive values at a later stage is not excluded, but depends on the details of the model, and specifically on the functional form of the branch $f_{\sscr\rm L}(\chi)$ corresponding to a large universe.

It is interesting to study the behaviour of $c_s^2$ in the models examined in Section~\ref{EffectiveGFT}, where a bouncing background is explicitly realised. To begin with, let us start from the special case $\alpha=0$, which reproduces the LQC effective dynamics for the cosmological background. The two branches $f_{\sscr\rm B}$, $f_{\sscr\rm L}$ in this case are given by Eqs.~(\ref{Eq:LQCbounceBranch}) and~(\ref{Eq:LQClateBranch}), respectively.
We find, using Eqs.~(\ref{SoundSpeed}) and~(\ref{DenSoundSpeedLQC})
\be\label{SoundSpeedLQC}
c_s^2=-\frac{1}{3}\left(1\pm \sqrt{1-q^2}\right)=-\frac{2}{3}\frac{\rho}{\rho_c} ~.
\ee
In the second step of (\ref{SoundSpeedLQC}), the upper sign corresponds to $f_{\sscr\rm B}$, whereas the lower one corresponds to $f_{\sscr\rm L}$.
We note that the speed of sound squared is always negative, has a minimum at the bounce $\left(c_s^2\right)_{\rm min}=-\frac{2}{3}$ when $\rho=\rho_c$, and approaches zero from below as $\rho\to0$. Given Eq.~(\ref{SoundSpeedLQC}), and recalling that maximal expansion rate in this model is reached at $\rho=\frac{\rho_c}{2}$, it is straightforward to check the general feature $c_s^2(\pm\chi_{\rm m})= -\frac{1}{3}$. We observe that $c_s^2$ is negative throughout cosmic history for the model with $\alpha=0$, and approaches zero from below in the large universe branch as $\chi$ tends to zero (cf.~Ref.~\cite{deHaro:2018sqw}). 
{It is interesting to compare these results with those obtained in Ref.~\cite{Cai:2012va} for a model based on generalised Galileons~\cite{Deffayet:2011gz}, where the speed of sound squared becomes negative---although only for a short period---around the bounce; see also Ref.~\cite{Cai:2014zga,Cai:2016hea} for a comparison between such effective models and the dynamics of perturbations in LQC.   In the models cited above gradient instabilities arise due to the violation of the null energy condition at the bounce (see also \cite{Libanov:2016kfc} and references therein). Recently, the possibility of establishing a theoretical no-go theorem regarding the realisation of a healthy non-singular bounce (i.e. free of pathologies such as gradient instabilities) has been discussed in the context of generalised Galileons, see Refs.~\cite{Kobayashi:2016xpl,Akama:2017jsa,Banerjee:2018svi}.}

The example examined above is just a particular case of the model reproducing the background dynamics of group field theory condensates, studied in Section~\ref{EffectiveGFT}. In the general case, i.e.~for $\alpha\neq0$, we have at the bounce
\be
\left(c_s^2\right)_{\rm min}=-\frac{2}{3}\left(1+\frac{\alpha}{V_{\sscr\rm B}}\right) ~.
\ee
In the large universe branch instead and for $\chi\simeq0$ we have, to leading order in $\chi$
\be
c_s^2\simeq\frac{3\alpha}{V_*}\sqrt{2+\frac{6\alpha}{V_*}}\,\frac{|\chi|}{\chi_{\rm m}} ~.
\ee
Thus, $c_s^2$ and $\alpha$ have the same sign in this regime. Therefore, for $\alpha<0$ the situation is qualitatively similar to the $\alpha=0$ case examined above, with a gradient instability extending also to the large universe branch.
For $\alpha>0$ the situation is different: there is a cross-over from $c_s^2<0$ near the bounce to $c_s^2>0$ when the universe is large. Such a cross-over must necessarily take place after the universe enters the phase of decelerated expansion, since $c_s^2= -\frac{1}{3}$ when $\dot{\chi}=0$ (see above). Thus, while the bounce is always accompanied by a gradient instability, the late universe branch would be characterised by a ghost instability for $\alpha>0$. We remark that the cross-over point where $c_s^2=0$ corresponds to a regime of strong coupling \cite{Ramazanov:2016xhp}.

\section{Discussion}\label{Conclusions}
We conclude by reviewing the main results obtained in this work and indicating directions for future studies.

In Section~\ref{Sec:Reconstruction} we illustrated in complete generality the theory reconstruction procedure for the function $f(\chi)$ in mimetic gravity. In the case of bouncing backgrounds, the implementation of the limiting curvature hypothesis requires that $f(\chi)$ be multivalued. This case was carefully examined and we gave general prescriptions to ensure continuity of $f(\chi)$ and its first derivative along the cosmic trajectory; in particular, by imposing suitable matching conditions at the branching points, both the energy density $\tilde{\rho}$ and pressure $\tilde{P}$ of the effective fluid are continuous throughout cosmic history. We showed that local properties of the function $f(\chi)$ are directly related to physically relevant quantities characterising the evolution of the cosmic background, such as the critical energy density and the bounce duration, as well as the equation of state of the effective fluid. In particular, the latter was shown to approach a constant value at late times, which is determined by the dominant matter species and the leading order term in the asymptotic expansion of $f(\chi)$ in that regime.

In Section~\ref{EffectiveGFT} we focused on a specific model obtained in Ref.~\cite{deCesare:2018cts}, where the function $f(\chi)$ was suitably reconstructed in order to reproduce the background evolution obtained from group field theory condensates in Ref.~\cite{\GFTbounce}. Quantities of physical interest were derived from local analysis of the two branches $f_{\sscr\rm B}$, $f_{\sscr\rm L}$, using the results of Section~\ref{Sec:Reconstruction}. The special case corresponding to the effective dynamics of LQC for a flat, isotropic universe was studied in detail. As an application, we studied the evolution of anisotropies near the bounce in a Bianchi~I universe for the model of Ref.~\cite{deCesare:2018cts}: our results generalise those obtained in Ref.~\cite{Chamseddine:2016uef} and show that anisotropies do not grow significantly during the bounce, and therefore do not spoil the smoothness of the bounce. It would be interesting to compare the results obtained in the effective approach considered here, with those of Ref.~\cite{deCesare:2017ynn}, where the dynamics of GFT condensates of anisotropic quanta was studied (see also Ref.~\cite{Pithis:2016cxg}). As discussed in Ref.~\cite{Bodendorfer:2017bjt,Bodendorfer:2018ptp}, the evolution of anisotropies is qualitatively similar in loop quantum cosmology and the corresponding mimetic gravity theory. It is therefore natural to ask whether an analogous statement can be made for GFT cosmology and the related model in mimetic gravity. We leave this question for future work. {Spherically symmetric geometries are also of interest and can be studied in the present framework by extending the analysis of Refs.~\cite{BenAchour:2017ivq,Chamseddine:2016ktu}.}

In Section~\ref{GravConst} we showed that there is an interesting reformulation of mimetic gravity involving two distinct time-varying effective gravitational constants $G_{\rm F}^{\sscr eff}$ and $G_{\rm R}^{\sscr eff}$, featuring respectively in the Friedmann and the Raychaudhuri equations. Consistency of such a description with the Bianchi identities is ensured by Eq.~(\ref{ConsistencyTwoGeff}), which is identically satisfied in mimetic gravity by all choices of the function $f(\chi)$. We derived the time evolution of the effective gravitational constants during the phase of decelerated expansion for $f(\chi)\sim \chi^p$, with $p>2$. We showed that the predicted time variation is too small to be observed if the limiting curvature hypothesis is realised. It would be of interest to further explore the consequences of the time variation of $G_{\rm F}^{\sscr eff}$ and $G_{\rm R}^{\sscr eff}$ in a more general and model independent setting.

Our discussion of perturbative instabilities in Section~\ref{Insta} highlights some serious limitations of bouncing models in mimetic gravity, which may hinder the possibility of using the simplest framework with the covariant action (\ref{ActionPrinciple}) for an effective description of quantum gravity in inhomogeneous spacetimes. The presence of gradient or ghost instabilities, which is a distinctive feature of mimetic gravity, seems to be even more serious in bouncing cosmologies; in fact, in such models the infinite age of the universe would offer no chance to keep instabilities under control.  {Remarkably, this issue has not been much appreciated in the literature on bouncing cosmologies in mimetic gravity. Based on the analogy with LQC (see Ref.~\cite{Cai:2014zga}), we expect the bounce to be accompanied by a short-lived gradient instability around the bounce affecting short-wavelength modes; however, there should be no instabilities away from the bounce.}
Some proposals to cure the instabilities by means of further modification of the mimetic gravity action have been made in Refs.~\cite{Gorji:2017cai,Hirano:2017zox}; however, their correspondence with the effective dynamics of quantum gravity models is yet to be established and shall be investigated in future work.


\acknowledgments{This work was partially supported by the Atlantic Association for Research in the Mathematical Sciences (AARMS) and by the Natural Sciences and Engineering Research Council of Canada (NSERC). It is a pleasure to thank Sabir Ramazanov and Edward Wilson-Ewing for helpful discussions on instabilities in mimetic gravity.}

\bibliographystyle{apsrev4-1M}
 \bibliography{references}

\end{document}